\documentclass[aps,prl,twocolumn,superscriptaddress,showpacs]{revtex4}

\usepackage{graphicx}

\begin{document}

\title{Imaging nano-scale electronic inhomogeneity in lightly doped Mott insulator Ca$_{2-x}$Na$_x$CuO$_2$Cl$_2$}

\newcommand{\Tokyo}{Department of Advanced Materials Science, University of Tokyo, Kashiwa-no-ha, Kashiwa, Chiba 277-8651, Japan}
\newcommand{\RIKEN}{RIKEN (The Institute of Physical and Chemical Research), Wako, Saitama 351-0198, Japan}
\newcommand{\JST}{Japan Science and Technology Corporation (JST), Kawaguchi, Saitama 332-0012, Japan}
\newcommand{\Kyoto}{Institute for Chemical Research, Kyoto University, Uji, Kyoto, 611-0011, Japan}

\author{Y. Kohsaka}
 \affiliation{\Tokyo}
\author{K. Iwaya}
 \affiliation{\RIKEN}
\author{S. Satow}
 \affiliation{\Tokyo}
\author{T. Hanaguri}
 \affiliation{\RIKEN}
 \affiliation{\JST}
\author{M. Azuma}
 \affiliation{\JST}
 \affiliation{\Kyoto}
\author{M. Takano}
 \affiliation{\Kyoto}
\author{H. Takagi}
 \affiliation{\Tokyo}
 \affiliation{\RIKEN}
 \affiliation{\JST}

\date{\today}

\begin{abstract}
The spatial variation of electronic states was imaged in the lightly doped Mott insulator Ca$_{2-x}$Na$_x$CuO$_2$Cl$_2$ using scanning tunneling microscopy / spectroscopy (STM/STS).
We observed nano-scale domains with a high local density of states within an insulating background.
The observed domains have a characteristic length scale of 2~nm ($\sim$4-5$a$, $a$ : lattice constant) with preferred orientations along the tetragonal [100] direction.
We argue that such spatially inhomogeneous electronic states are inherent to slightly doped Mott insulators and play an important role for the insulator to metal transition.
\end{abstract}


\maketitle

Holes and electrons doped into antiferromagnetic Mott insulators can bring about many exotic phenomena such as high-temperature superconductivity (HTS) in copper oxides and colossal magnetoresistance (CMR) in manganese oxides~\cite{Imada98}.
When Mott insulators are doped lightly, a conflict arises between the itinerancy of the doped charge carriers and the energy gain of the magnetic background, which can cause inherently spatially inhomogeneous electronic states, such as stripes~\cite{Zaanen89,Emery90,Grilli91,Emery93,Kivelson98}.
It has been suggested that such an inhomogeneity is a key ingredient of HTS and CMR~\cite{Kivelson98,Burgy01,Burgy03}.

Indeed, spatially inhomogeneous superconductivity has been observed in Bi$_2$Sr$_2$CaCu$_2$O$_y$ by scanning tunneling microscopy / spectroscopy (STM/STS)~\cite{Pan01,Howald01,Lang02}.
It is not yet clear however whether or not the observed inhomogeneity is electronic in origin or related to the local off-stoichiometry.
This is, in our opinion, because the materials used in these STM studies were heavily doped superconductors, whereas purely the electronic phase separation is anticipated to occur near the doping-induced insulator to metal transition (IMT) found in the lightly doped region.
It is therefore highly desirable to extend such real-space imaging into very lightly doped region.
Bi$_2$Sr$_2$CaCu$_2$O$_y$, however, is not chemically stable when lightly doped and is therefore an unsuitable choice of material to extend STM/STS studies towards the IMT.

\begin{figure}
	\includegraphics{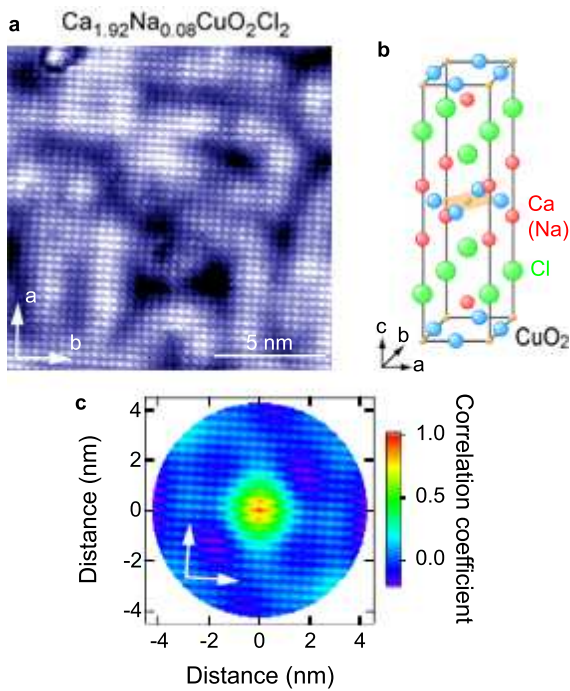}
	\caption{
		A 15~nm $\times$ 15~nm STM images of (001) cleavage planes of Ca$_{1.92}$Na$_{0.08}$CuO$_2$Cl$_2$ single crystals taken at a sample bias ($V_{\rm s}$) of 200~mV, a tunneling current ($I_{\rm t}$) of 10~pA and at 7.0~K.
		(b) Crystal structure of Ca$_{2-x}$Na$_x$CuO$_2$Cl$_2$.
		(c) Auto-correlation patterns of (a).
	\label{fig:STM}}
\end{figure}

An alternative candidate with which to explore the real space physics of lightly doped Mott insulators is the cupric oxychloride Ca$_{2-x}$Na$_x$CuO$_2$Cl$_2$.
The parent compound Ca$_2$CuO$_2$Cl$_2$ is an antiferromagnetic Mott insulator.
By substituting Na for Ca, the system is doped with holes and eventually develops into a superconductor at $x\sim 0.08$~\cite{Hiroi9496}.
Moreover, single crystals of Ca$_{2-x}$Na$_x$CuO$_2$Cl$_2$ are very easy to cleave, making them ideal systems for STM/STS studies.
As shown in Fig.~\ref{fig:STM}b, Ca$_{2-x}$Na$_x$CuO$_2$Cl$_2$ has a crystal structure closely related to those of La$_{2-x}$Sr$_x$CuO$_4$ but maintains undistorted tetragonal CuO$_2$ layers down to low temperatures.
Such a simple structure with small number of constitution elements minimize the possibility of the local off-stoichiometry other than doping.
In this Letter, we report atomic resolution STM/STS results on Ca$_{2-x}$Na$_x$CuO$_2$Cl$_2$ with various hole concentrations across a doping-induced IMT.
A variation of electronic states, in the form of nano-scale domains with spatially preferred orientations, was clearly resolved.
We suggest that the observed electronic inhomogeneity is inherent to slightly doped Mott insulators.

Sizeable single crystals of Ca$_{2-x}$Na$_x$CuO$_2$Cl$_2$ were grown using a novel high-pressure flux technique over a wide concentration range across the IMT boundary~\cite{Kohsaka02}.
We have carried out STM/STS measurements on single crystals from $x = 0.06$ (insulator) to $x = 0.10$ (heavily underdoped superconductor with $T_{\rm c} = 13$~K) under ultrahigh vacuum (UHV) conditions of $\sim 10^{-8}$~Pa.
The samples were cleaved along (001) planes in UHV at about 80~K and immediately transferred to the STM head and cooled to 4.7-7.0~K, where all reported data were taken.
The exposed surface layer is believed to be the highly ionic Ca-Cl layer, since bonding to the adjacent Ca-Cl layer is weakest at this interface.

A typical constant-current STM image of $x = 0.08$ crystal, which is in the critical vicinity of the IMT boundary, is shown in Fig.~\ref{fig:STM}a.
Regular square atomic lattices are clearly observed, and the lattice spacing is estimated to be 0.39~nm, in good agreement with the in-plane lattice constant of the tetragonal unit cell~\cite{Kohsaka02}.
Superimposed on the regular atomic lattices, we observe nano-scale corrugations, which can be seen as the bright (high) and dark (low) regions in Fig.~\ref{fig:STM}a.
The corrugations are not spatially regular but rather show a preferred orientation with respect to a specific crystallographic axis.
As shown in Fig.~\ref{fig:STM}a, the domains in many cases are aligned along [100] or [010], and form river-like pathways.
Some images appear more patchy, but again the patches are aligned along [100] or [010].
Auto-correlation analyses shown in Fig.~\ref{fig:STM}c clearly and quantitatively demonstrate an enhanced spatial correlation along the [100] and [010] directions.
Around the origin, the correlation shows a rapid decay on the length scale of $\sim$2~nm, which corresponds to the width of the rivers or the size of the patches.
A weak tail then follows, representing the river-like spatial correlation.

\begin{figure}
	\includegraphics{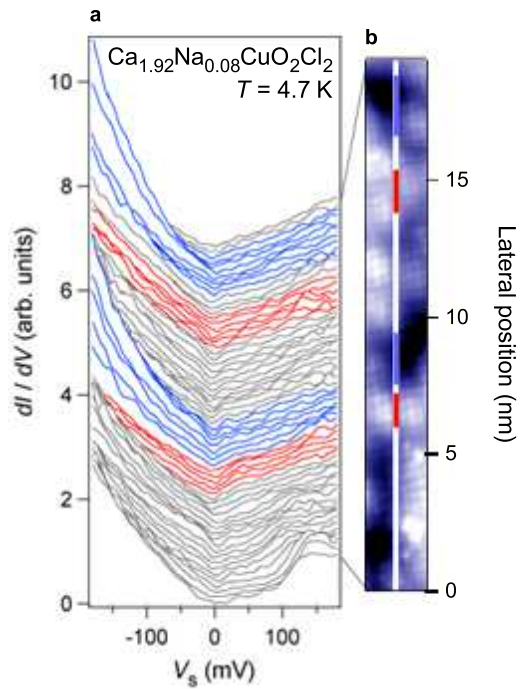}
	\caption{
		(a) $dI/dV$ spectra measured along the line shown in (b).
		Each spectrum is shifted vertically dependent on the position.
		For clarity, the spectra taken at apparently lower (higher) positions are colored blue (red).
		(b) STM image with the trajectory along which the spectra in (a) were measured.
		The colored-line code corresponds to the positions of the spectra individually-colored shown in (a).
	\label{fig:STS}}
\end{figure}

We ascribe the observed corrugations to a spatial variation of electronic states, rather than a topographic effect.
It is emphasized here that the height change, detected by constant-current imaging, represents not only the $z$-axis coordinates of the surface atoms but also the spatial variations of the local density of states (LDOS).
If there exists a region with a lower LDOS than its surroundings, the STM tip will come close to the surface in order to keep the tunneling current constant.
During our STM measurements, we found that the magnitude of the corrugations strongly depends on the bias voltage.
This implies that the observed corrugations cannot be understood simply in terms of a topographic effect, but rather by a spatial variation of the energy dependence of the LDOS.
Such spatial variation in LDOS was indeed confirmed by the STS measurements shown in Fig.~\ref{fig:STS}a.
By comparing Fig.~\ref{fig:STS}a and Fig.~\ref{fig:STS}b, it is clear that there is a strong correlation between the shape of the differential conductance ($dI/dV$) spectra, which is proportional to the LDOS, and the apparent height.
This demonstrates that the electronic states of underdoped Ca$_{2-x}$Na$_x$CuO$_2$Cl$_2$ are spatially inhomogeneous and suggests the formation of distinct domains in the local electronic structure.
It may be interesting to infer that the length scale of the domains observed here is comparable to the size of superconducting islands observed in heavily doped Bi$_2$Sr$_2$CaCu$_2$O$_y$~\cite{Pan01,Howald01,Lang02}.

\begin{figure}
	\includegraphics{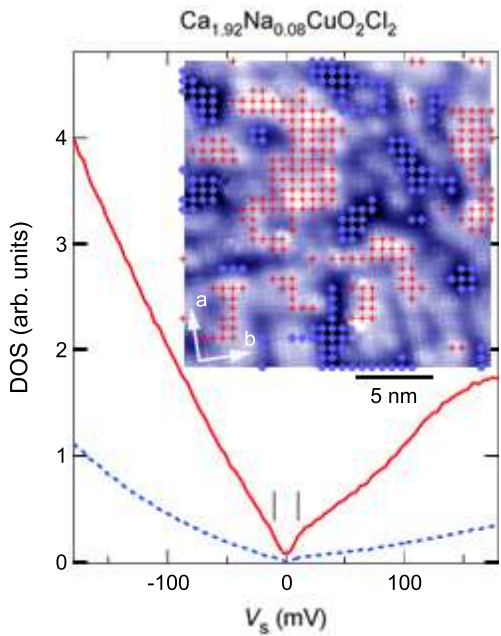}
	\caption{
		The averaged LDOS from the high (red, solid line) and the low (blue, broken line) domains are shown for a $x = 0.08$ sample. 
		The LDOS was estimated from the $dI/dV$ spectrum by calibrating the tip-sample distance.
		The vertical ticks located $\pm$ 10~meV indicate the knees in the LDOS spectra.
		The inset shows the topographic image taken at $V_{\rm s} = 200$~mV and $I_{\rm t} \sim 100$~pA with markers where the spectra were taken (red cross=bright / blue diamond=dark).
		Colors of markers correspond to the colors of spectra in the main panel.
	\label{fig:LDOS}}
\end{figure}

The $dI/dV$ spectra in Fig.~\ref{fig:STS}a do not represent the {\it magnitude} of the LDOS, due to the change in the tip-sample distance in constant current imaging.
However, if the tunneling barrier is uniform, we can easily estimate a correction associated with the change in the tip-sample distance $z$, since $dI/dV = C\exp(-2\kappa z)LDOS$, where $C$ is a constant and $\kappa$ is related to a tunneling barrier~\cite{Tersoff8385}.
The tunneling barrier height was measured along the line shown in Fig.~\ref{fig:STS}b and was found to be uniformly $3.3 \pm 0.2$~eV.
We thus performed a numerical correction of the tip-sample distance and estimated the relative magnitude of the LDOS at each position.
Fig.~\ref{fig:LDOS} indicates the comparison of the relative LDOS averaged over the bright and dark regions in the inset of Fig.~\ref{fig:LDOS}
It is evident from this plot that LDOS is substantially suppressed in the dark regions up to a few hundred meV.
Therefore, the dark region is substantially less metallic (more insulating) than the bright region.
We ascribe the V-shaped behavior of LDOS to the large pseudogap ($\sim$0.1-0.2~eV) universally observed in underdoped cuprates by angle-resolved photoemission spectroscopy (ARPES) near $(\pi,0)$.
Since this large pseudogap becomes larger upon approaching the undoped Mott insulator~\cite{Ronning98}, the gap enhancement in the dark regions implies the suppression of hole density.
Besides the large pseudogap, a knee-like structure is found around $\pm$10~meV at the bottom of the V-shape.
We suspect that this might originate from superconductivity.

As the Ca and Cl atoms are strongly ionic, their electronic states are energetically far away from the Fermi level.
We therefore believe that the majority of the tunneling current observed originates from the uppermost CuO$_2$ plane just below the exposed Ca-Cl layer, and that the electronic inhomogeneity takes place within the CuO$_2$ plane.
It might be argued that the observed phenomena are surface rather than bulk effects.
However, this system is extremely two-dimensional in nature and the coupling between the adjacent CuO$_2$ planes is negligibly small.
It is therefore likely that the electronic states of the topmost CuO$_2$ layer, sandwiched by ionic Ca-Cl layers, resemble those in the bulk.
Recent transport measurements indicate that the anisotropy between the in-plane and the out-of-plane resistivities reaches 10$^4$ even in the superconducting $x = 0.10$ composition~\cite{Waku}.
This is one of the largest anisotropic resistivity ratios ever observed in the cuprate superconductors.

We consider two scenarios for the origin of observed electronic inhomogeneity within the CuO$_2$ plane.
One is a poorly screened impurity potential.
The dopant Na${\rm ^+}$, substituted for Ca${\rm ^{2+}}$, should modify the electrostatic potential of the neighboring CuO$_2$ planes, which can give rise to inhomogeneous distribution of carriers.
In ordinary metals, however, the electrostatic potential is screened by itinerant carriers over a subatomic spacing~\cite{AshcroftMermin} and can not simply lead to a nanoscale inhomogeneity.
This situation may be altered significantly in lightly doped cuprates.
A close proximity to the Mott insulator can bring about a nonlinear screening effect~\cite{Wang01,Wang02} or a reduction of the itinerant coherent states~\cite{Boeri03}.
As a result, effective charge screening length can be anomalously long and a {\it nano-scale} electronic inhomogeneity may emerge.

Another possible scenario for the observed inhomogeneity may be the electronic phase separation.
Electronic phase separation arising from competing ordered states is at the heart of the physics of CMR phenomena in the manganites~\cite{Burgy01,Burgy03}, where nano-scale phase separation between a ferromagnetic metal and a charge ordered insulator has been observed by STM/STS~\cite{Fath99,Renner02,Becker02}.
Another strong case for electronic phase separation may be made for the lightly doped cuprates, and this has been discussed to be intimately connected to the mechanism of high temperature superconductivity~\cite{Kivelson98}.
If this is the case, naively, two distinct phases should be observed and the ratio of the two should change upon doping.

\begin{figure}
	\includegraphics{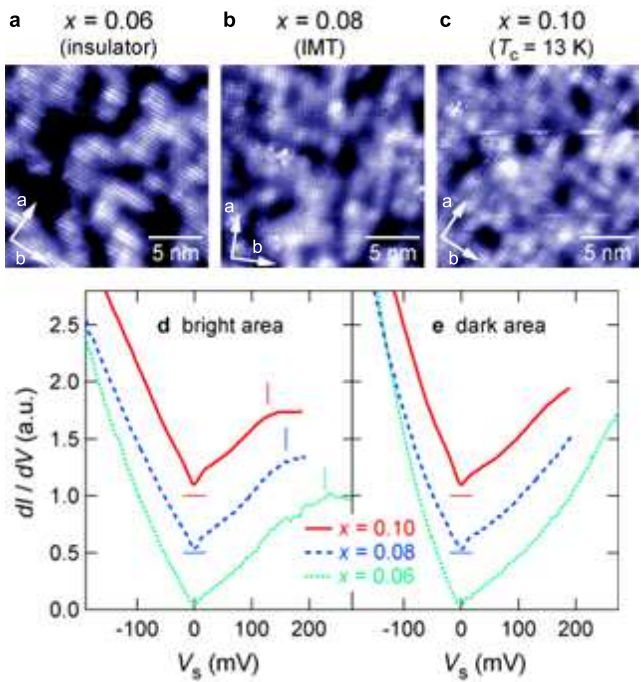}
	\caption{
		Doping dependence of STM images and $dI/dV$ spectra of Ca$_{2-x}$Na$_x$CuO$_2$Cl$_2$.
		The topographic images (a), (b), and (c) were from $x$ = 0.06 (insulator), $x$ = 0.08 (right at IMT), and $x$ = 0.10 (superconductor) samples, respectively.
		(a) was taken at $V_{\rm s}$ = 300~mV, $I_{\rm t} \sim$ 100~pA, and $T$ = 5.0~K.
		(b) and (c) were taken at  $V_{\rm s}$ = 200~mV, $I_{\rm t} \sim$ 100~pA, and $T$ = 4.7~K.
		The images from $x$ = 0.08 (b) was taken on different surface from that of Fig.~\ref{fig:LDOS}, demonstrating the reproducibility of main features.
		(d) and (e) are averaged differential conductance spectra collected from the bright and dark regions, respectively.
		The spectra are shifted vertically for clarity and the horizontal bars indicate zero.
		The vertical tick marks indicate the energies of the shoulder structures in the spectra.
	\label{fig:xdep}}
\end{figure}

To examine this point, we performed STM/STS on the samples with different dopings and found systematic evolution of both STM images and $dI/dV$ spectra as shown in Fig.~\ref{fig:xdep}.
It is clear as shown in Fig.~\ref{fig:xdep}a-c that the nano-scale mixture of high and low LDOS regions is observed regardless of the ground state, insulator or superconductor.
The high LDOS (bright) area increases with increasing density of holes, and with further doping, it is likely that the electronic inhomogeneity will eventually fade out.
The system appears to change from an insulator to a metal once the metallic domains become connected over the entire surface.
Although these appear to be consistent with the phase separation, the evolution of the electronic states within the two regions can not be understood in its naivest form.
As shown in Fig.~\ref{fig:xdep}d and Fig.~\ref{fig:xdep}e, while the shape of $dI/dV$ spectra in the dark regions are nearly identical, those in the bright regions show a substantial change as a function of doping, which is not consistent with ``naive'' phase separation picture.
Even if a kind of electronic phase separation is responsible for the observed phenomena, additional ingredient, for example, poorly-screened impurity potential, should be invoked.
In any event, we believe that the observed nano-scale electronic inhomogeneity represents the physics of the lightly doped Mott insulator.

In conclusion, STM/STS measurements on underdoped Ca$_{2-x}$Na$_x$CuO$_2$Cl$_2$ revealed pronounced electronic inhomogeneity near the IMT.
The observed features of inhomogeneity, i.e. a long characteristic length scale of $\sim$2~nm and the evolution of the LDOS as a function of doping, can not be ascribed to the conventional charge screening nor to the naive electronic phase separation.
We infer that competing orders inherent to doped Mott insulators plays an important role for the observed electronic inhomogeneity.
Regardless the origin, real space physics of lightly doped Mott insulator is indispensable in understanding the insulator to metal transition in a real system.

We thank N. E. Hussey, N. Shannon, and K. M. Shen for critical reading of manuscript.
This work was partly supported by a Grant-in-Aid for Scientific Research from the MEXT, Japan.

{\it note added in proof:}
Spatial modulations of tunneling conductance with checkerboard-like patterns have recently observed in the same compound.
This is to be published elsewhere.

\bibliographystyle{apsrev}
\bibliography{condmat_Kohsaka}

\end{document}